\documentclass[conference]{IEEEtran}
\IEEEoverridecommandlockouts
\usepackage{cite}
\usepackage{amsmath,amssymb,amsfonts}
\usepackage{booktabs}
\usepackage{graphicx}
\usepackage{textcomp}
\usepackage{xcolor}
\usepackage{algorithm}
\usepackage{algpseudocode}
\usepackage{float}
\usepackage{enumitem}
\usepackage{multirow}
\usepackage{graphicx}
\usepackage{array} 
\usepackage{tcolorbox}
\usepackage{orcidlink}
\usepackage{threeparttable, threeparttablex}

\makeatletter
\newcommand{\linebreakand}{%
  \end{@IEEEauthorhalign}
  \hfill\mbox{}\par
  \mbox{}\hfill\begin{@IEEEauthorhalign}
}
\makeatother

\begin{document}


\title{The Privacy and Security Risks of Edge LLM Optimization}






\author{\IEEEauthorblockN{Mingchen Li\,\orcidlink{0000-0003-4553-9598}}
\IEEEauthorblockA{\textit{University of South Florida}\\
Tampa, FL, USA}
\and
\IEEEauthorblockN{Ren-Yi Huang\,\orcidlink{0009-0006-8341-5835}}
\IEEEauthorblockA{\textit{University of South Florida}\\
Tampa, FL, USA}
\and
\IEEEauthorblockN{Dumindu Samaraweera\,\orcidlink{0000-0003-4097-5585}}
\IEEEauthorblockA{\textit{Embry-Riddle Aeronautical University}\\
Daytona Beach, FL, USA}
\linebreakand
\IEEEauthorblockN{Prashant Shekhar\,\orcidlink{0000-0003-2353-6740}}
\IEEEauthorblockA{\textit{Embry-Riddle Aeronautical University}\\
Daytona Beach, FL, USA}
\and
\IEEEauthorblockN{J. Morris Chang\,\orcidlink{0000-0002-0660-7191}}
\IEEEauthorblockA{\textit{University of South Florida}\\
Tampa, FL, USA}
}

\maketitle

\begin{abstract}
    \looseness-1Edge deployment of Large Language Models requires optimization techniques that reduce memory, computation, and energy costs. However, preserving task performance does not necessarily preserve security. We argue that optimization is an information transformation that can alter representations, decision pathways, and trust boundaries, creating security drift and optimization-induced attack surfaces.
\end{abstract}

\section{Introduction}
Large Language Models (LLMs) are rapidly moving beyond cloud-hosted services to edge platforms such as enterprise workstations, autonomous vehicles, industrial gateways, mobile devices, and tactical computing systems. Deploying intelligence closer to data sources offers compelling advantages, including reduced latency, improved privacy, greater operational resilience, and compliance with increasingly stringent data sovereignty requirements. These benefits have accelerated the adoption of edge AI across safety-critical domains, including healthcare, manufacturing, transportation, defense, and critical infrastructure.


However, edge platforms operate under strict memory, compute, and energy budgets. While quantization, pruning, model partitioning, parameter-efficient fine-tuning (PEFT), and emerging architectures (e.g., MoE, SSMs) make deployment feasible, they challenge a core AI security assumption: that an optimized model remains security-equivalent to its uncompressed baseline. The problem is that optimization does not simply make a model smaller or faster. It changes \emph{how information is represented, processed, distributed, or exposed}. For instance, quantization changes numerical precision; pruning removes computational pathways; PEFT concentrates adaptation into a small set of parameters; and model partitioning moves intermediate representations across trust boundaries. Architectural approaches can similarly change routing, state representation, and execution pathways. Thus, an optimization that preserves average task utility can nevertheless alter properties that matter for security, including robustness margins, refusal behavior, activation separability, fault sensitivity, or the amount of information available to an observer.

We therefore argue that edge optimization should be viewed as an information transformation, rather than merely an engineering optimization. Let ($\mathcal{T}$) represent a deployment operator that transforms an original model into an optimized representation. Conceptually, the process can be viewed as

\[
\begin{aligned}
&\text{Original model information} \\
&\quad \xrightarrow{\mathcal{T}} \text{Distorted or redistributed} \\
&\quad\phantom{\xrightarrow{\mathcal{O}}} \text{representation} \\
&\quad \rightarrow \text{Changed decision geometry or} \\
&\quad\phantom{\rightarrow} \text{information exposure} \\
&\quad \rightarrow \text{Security drift}.
\end{aligned}
\]

The key observation is that optimization is usually guided by a task- or systems-oriented objective: preserve accuracy, reduce latency, minimize memory, or lower energy consumption. The information needed to satisfy these objectives, however, is not necessarily the same information that supports security. Compression and representation-learning perspectives provide a useful theoretical basis for this distinction: an optimization process selectively preserves information according to its objective and discards, restructures, or concentrates other information. Consequently, small changes in task utility do not imply equally small changes in security properties.

We term this phenomenon \textbf{security drift}: a shift in a model's security characteristics caused by an efficiency transformation. Security drift does not mean optimization always degrades safety, but rather that it alters the conditions under which security properties hold. An optimization may preserve benchmark accuracy while shifting responses near safety boundaries, increasing perturbation sensitivity, or exposing intermediate computations. Security must therefore be evaluated on the deployed representation. This perspective fundamentally expands the attack surface of edge LLMs beyond conventional threats like malicious prompts or API exploitation. Optimized deployments introduce new security-relevant assets: low-precision states, sparse structures, adaptation modules, intermediate activations, dynamic routing choices, and hardware execution pathways. Attackers exploit these properties because optimization changes where information resides, how it flows, and what becomes observable. We call these deployment-conditioned vulnerabilities \textbf{optimization-induced attack surfaces (OIAS)}. Whether through numerical distortion, parameter removal, information concentration, or activation exposure, OIAS stem from a single origin: efficiency transformations reshape the model's security-relevant information structure.

This article examines this relationship across widely used and emerging approaches to efficient edge LLM deployment. We first establish the information-distortion perspective and explain how optimization can produce security drift. We then examine optimization-induced attack surfaces associated with quantization, pruning, model partitioning, and PEFT, followed by emerging security considerations in efficient architectures such as MoE and SSM-based models. Finally, with a case study, we discuss security-aware optimization and argue for deployment workflows that evaluate efficiency and security jointly rather than treating security as a post-optimization validation step. Our central message is simple: \emph{making an LLM smaller, faster, or more efficient can also change what information the system retains, where that information resides, and what an adversary can observe or manipulate. Trustworthy edge AI therefore requires optimizing not only for efficiency, but also for the preservation of security properties}.

\section{Why Optimization Changes the Threat Model}
Optimization is usually judged by model size, latency, throughput, memory, and energy, whereas security is often evaluated on the original model or only after optimization. This workflow implicitly assumes that security properties transfer from the original model to its optimized deployment. For edge LLMs, that assumption is unsafe: quantization changes numerical states, pruning changes active pathways, PEFT creates portable adaptation modules, and partitioning moves intermediate representations across trust boundaries. The threat model must therefore include deployment-specific assets and execution conditions, not only base parameters and inputs.

\begin{figure*}
    \centerline{\includegraphics[width=34pc]{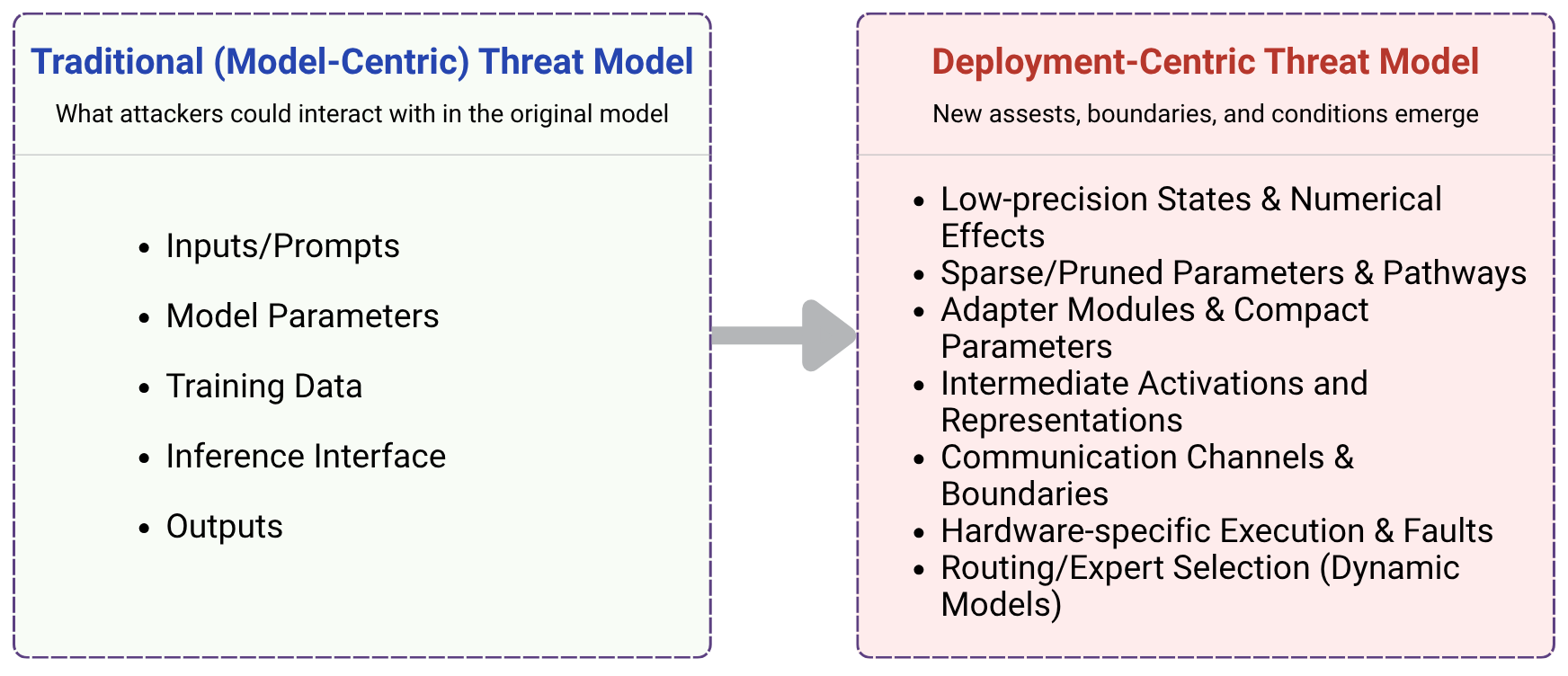}}
    \caption{Edge optimization expands the security boundary beyond conventional model-centric targets to include low-precision states, sparse pathways, adaptation modules, intermediate representations, communication channels, hardware-specific execution, and dynamic routing.}
    \label{fig:threat_model}
    \vspace*{-5pt}
\end{figure*}

\subsection{From Model-Centric to Deployment-Centric Threats}

Conventional model-centric threat models focus on inputs, base parameters, training data, or inference APIs. In contrast, optimized edge deployments expand this boundary to include intermediate activations, low-precision states, adapter modules, communication channels, hardware-specific execution, and dynamic routing decisions (Fig. \ref{fig:threat_model}).

Crucially, many optimization-induced attacks leverage access vectors non-existent in full-precision baselines. Split inference exposes intermediate activations across network boundaries; low-precision execution induces sensitivity to hardware bit-flips; and modular adapters introduce provenance and tampering risks due to independent distribution. Threat models are thus strictly deployment-specific: two deployments of the same foundation model can exhibit fundamentally distinct attack surfaces based on their precision, sparsity, partitioning, and hardware execution environment.

This observation motivates the central question of the next section: \emph{what happens to security when the information structure of a model is transformed to meet deployment constraints?} We use the concept of \textbf{information distortion and security drift} to explain this relationship and to connect different optimization techniques to the attack surfaces they create.

\section{Information Distortion and Security Drift}
To understand the relationship between optimization techniques and the attack surfaces they introduce, we first examine how optimization transforms the model and its deployment. Edge deployment transforms the original model parameters $\theta$ into an optimized representation $\hat{\theta}$ through an efficiency-driven operator,
\begin{equation*}
\hat{\theta} = \mathcal{T}{\eta}(\theta),
\end{equation*}
where $\mathcal{T}{\eta}$ represents an optimization operation controlled by parameters $\eta$, such as quantization bit-width, pruning sparsity, adaptation rank, or partitioning configuration. Conventional optimization primarily asks whether task performance is preserved:
\begin{equation*}
\mathcal{L}(f_{\theta}) \approx \mathcal{L}(f_{\hat{\theta}}).
\end{equation*}
However, this condition does not imply that security-relevant properties are preserved. To make this distinction explicit, let $Z_l$ and $\hat{Z}_l$ denote the intermediate representation at layer $l$ before and after optimization. In general,
\begin{equation*}
Z_l \neq \hat{Z}l,
\end{equation*}
even when the two models exhibit nearly identical behavior on standard task benchmarks. We refer to this change as \emph{information distortion}: the alteration of internal representations, computational pathways, or information exposure induced by an efficiency transformation. The magnitude and nature of this distortion depend on $\mathcal{T}{\eta}$ and need not be aligned with the task loss. Thus, an optimization can preserve information sufficient for common task predictions while altering information or structures that support robustness, safety, privacy, or integrity.

This distinction motivates our notion of \textbf{security drift}, which captures the change in a security property between the original and optimized deployments:
\begin{equation*}
\Delta \mathcal{S}
=
\left|
\mathcal{S}(f_{\theta}) -
\mathcal{S}(f_{\hat{\theta}})
\right|,
\end{equation*}
where $\mathcal{S}(\cdot)$ denotes a security characteristic, such as jailbreak resistance, adversarial robustness, privacy leakage, or extraction resilience. Importantly, $\Delta\mathcal{S}$ need not have a fixed sign: optimization may weaken, preserve, or in some cases improve a security property. Security drift therefore describes \emph{change} rather than necessarily degradation. The key theoretical observation is that
\begin{equation*}
\mathcal{L}(f_{\theta}) \approx \mathcal{L}(f_{\hat{\theta}})
;\not\Rightarrow;
\Delta\mathcal{S} \approx 0.
\end{equation*}
This decoupling between task preservation and security preservation provides the foundation for the security-efficiency paradox. With that understanding, optimization alters security-relevant information through four primary mechanisms:

\begin{itemize}
    \item \textbf{Numerical Distortion:} Low-precision discretization perturbs activation values and decision geometry ($D_l = d(Z_l, \hat{Z}_l)$). Small aggregate task loss changes can disproportionately destabilize inputs near safety-critical or fault-sensitive decision margins.   
    
    \item \textbf{Structural Distortion:} Pruning eliminates parameters and computational pathways. Because task-oriented weight importance rarely coincides with security-oriented importance, removing 'redundant' parameters can destroy critical safety or robustness pathways.   
    
    \item \textbf{Information Concentration:} PEFT confines adaptation updates to compact parameter subspaces. Concentrating behavioral knowledge into portable modules increases their vulnerability to extraction, stealing, and localized tampering or privacy leakage \cite{li2025compleak}.   
    
    \item \textbf{Information Exposure:} Partitioned inference transmits intermediate activations ($Z_l$) across hardware, network, or organizational trust boundaries ($\mathcal{E}=\mathcal{E}(\hat{\theta},\mathcal{T}_{\eta})$), exposing internal representations to interception and input reconstruction.
\end{itemize}

\begin{figure*}
    \centerline{\includegraphics[width=42pc]{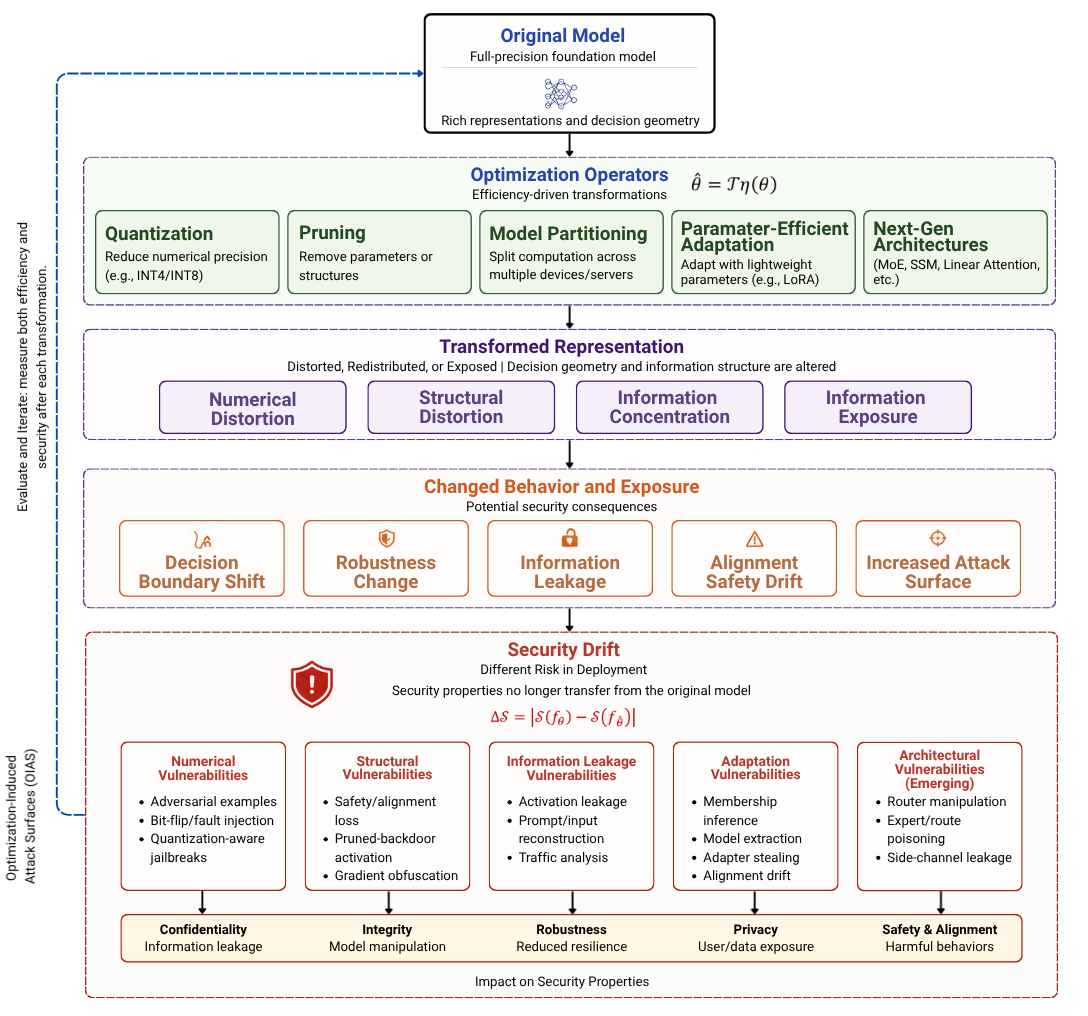}}
    \caption{Overview of optimization-induced security drift. Edge deployment constraints necessitate optimization operators (e.g., quantization, pruning, PEFT, and model partitioning), which alter the model's internal information representations. These changes propagate through four information distortion mechanisms, ultimately producing measurable changes in security properties and expanding the attack surface.}
    \label{fig:security_drift}
    \vspace*{-5pt}
\end{figure*}

These mechanisms provide a common explanation for the optimization-induced attack surfaces examined next and are summarized in Fig.~\ref{fig:security_drift}. Although quantization, pruning, PEFT, and partitioning operate at different levels, each can modify a security-relevant aspect of the deployed system through numerical distortion, structural distortion, information concentration, or information exposure. Security evaluation must therefore ask not only whether an optimization preserves what the model can do, but also whether it preserves the conditions under which the model remains secure.

\section{Optimization-Induced Attack Surfaces}
The preceding framework provides a common lens for understanding how efficiency transformations can alter security. However, the effects depend on what an optimization changes in the deployed system. The following four cases illustrate this distinction: quantization changes numerical representation, pruning changes computational structure, PEFT concentrates adaptation information, and model partitioning changes where intermediate information is exposed. Moreover, these security changes can also become potential consequential when optimized LLMs are deployed in agentic systems, where model outputs may directly influence downstream decisions and actions.

\subsection{Quantization: Trading Numerical Precision for New Vulnerabilities}


Quantization reduces memory and energy footprint by representing weights and activations in lower precision (e.g., INT8, INT4). Modern post-training methods (e.g., AWQ, GPTQ) preserve benchmark accuracy well \cite{lin2024awq}. However, from a security standpoint, quantization fundamentally changes an LLM's numerical behavior. Discretization alters activation distributions, compresses dynamic range, and shifts decision boundaries. While these shifts rarely degrade benchmark utility, they alter responses to perturbations, hardware faults, and adversarial prompts. A quantized model is thus not merely a smaller model, but a computationally distinct system with unique security risks.

One consequence is \emph{increased sensitivity to hardware-level perturbations}. Because low-precision representations use fewer bits, individual bit errors caused by hardware faults, memory corruption, or deliberate fault injection can have a disproportionate effect on model behavior. An attacker capable of inducing bit flips in memory or accelerator hardware may manipulate inference outcomes, bypass safety mechanisms, or trigger unexpected responses \cite{egashira2024exploiting, guo2026tfl}. This risk is particularly relevant for edge deployments that rely on heterogeneous accelerators operating in less controlled environments than cloud data centers.

Quantization may also \emph{influence adversarial robustness}. Numerical changes introduced during optimization can alter the geometry of the model's decision space and affect how adversarial prompts and perturbations propagate through the network. The resulting security impact can depend on the quantization strategy, calibration procedure, and deployment configuration. Consequently, accuracy and perplexity alone provide limited insight into resilience against malicious inputs. In agentic systems, these effects can extend beyond generated text to tool selection and action execution. Recent evaluations show that quantization can degrade tool-use and broader agentic performance even when overall LLM capabilities remain comparatively stable \cite{dong2025compressed}. Thus, modest precision changes at the model level may become more consequential when outputs directly govern downstream actions.

Another concern is the preservation of \emph{safety alignment}. Alignment techniques are typically developed and validated using high-precision models, but quantization can alter parameter values and activation patterns in ways that weaken learned safety behaviors. A model that satisfies alignment benchmarks before optimization may therefore respond differently to jailbreak prompts or safety-critical queries after deployment \cite{egashira2025mind}. Recent work specifically shows that quantization errors can contribute to safety-alignment degradation \cite{yangtoward}. As examined in our later case study, this observation suggests that safety properties should be re-evaluated after optimization rather than assumed to transfer unchanged from the original model.

Mitigating these risks requires extending conventional quantization workflows with security-aware evaluation. In addition to accuracy, latency, and memory reduction, optimized models should be assessed for adversarial robustness, fault tolerance, privacy preservation, and alignment stability. Security-aware calibration, post-quantization red teaming, and hardware-assisted integrity protections can help identify vulnerabilities before deployment. More broadly, quantization should be treated as a security-relevant transformation requiring systematic validation throughout the optimization pipeline.

\textit{Security Takeaway: Quantization improves computational efficiency by changing numerical representation, but security properties that depend on numerical sensitivity must be revalidated after deployment.}

\subsection{Pruning: Structural Changes, Hidden Dependencies}



Pruning reduces compute and memory overhead by stripping parameters, attention heads, or structural layers deemed redundant for task accuracy. Unlike quantization, pruning alters the model's physical graph topology, removing computational pathways and structural redundancy crucial for adversarial robustness and safety reasoning. A critical vulnerability is the \emph{unintended elimination of safety subnetworks}. Safety alignment is localized within specific attention heads and parameter groups. Because standard pruning algorithms optimize purely for loss minimization, they indiscriminately excise lower-magnitude safety weights, degrading refusal behavior and jailbreak resistance \cite{egashira2025fewer} while leaving standard benchmark accuracy intact.

Pruning may also reduce \emph{adversarial robustness} by removing redundant computational pathways that provide resilience to unexpected or carefully crafted inputs. This can make model behavior more brittle and narrow the internal representations available during inference \cite{huang2025antidote}. Such structural changes can also complicate security auditing when pruning disrupts neurons or attention heads associated with interpretable linguistic or safety-related functions.

These effects become particularly important in \emph{agentic systems}, where model outputs influence planning, tool use, and downstream actions. Empirical evaluations show that pruning can degrade tool-use and broader agent capabilities \cite{dong2025compressed}, potentially propagating structural changes into consequential agent decisions.

Mitigating these risks requires moving beyond pruning objectives based solely on weight magnitude or computational contribution. Security-aware pruning should preserve robustness, alignment, and critical reasoning capabilities, followed by post-pruning evaluation of jailbreak resistance, adversarial robustness, and safety behavior. This ensures that computational efficiency does not come at the cost of disproportionately increasing the attack surface of edge-deployed LLMs.

\textit{Security Takeaway: Pruning can preserve task performance while removing structural redundancy or components that contribute disproportionately to security.}

\subsection{Model Partitioning: When Efficiency Expands the Trust Boundary}
Model partitioning enables LLM inference when a single edge device cannot accommodate the model's computational and memory requirements. Instead of executing the entire model on one platform, inference is divided across multiple devices or between edge and cloud resources, balancing latency, bandwidth, and energy consumption.

Unlike quantization and pruning, which modify the model's numerical representation or computational structure, partitioning changes the \emph{execution environment}. Intermediate activations, hidden representations, and attention states must cross communication boundaries that do not exist in monolithic deployments. These representations can contain rich semantic information and reveal more about user inputs than expected \cite{luo2025prompt}.

A primary concern is \emph{activation leakage}. Although intermediate representations are often treated as transient computational artifacts, they can encode substantial information about original inputs. An adversary who gains access to these representations may reconstruct prompts, infer sensitive attributes, or recover task-specific information. This creates confidentiality risks even when raw inputs remain encrypted or inaccessible, particularly in privacy-sensitive edge applications.

Partitioning also introduces \emph{additional trust boundaries}. Communication channels between edge devices, accelerators, and cloud services become targets for eavesdropping, manipulation, replay, and man-in-the-middle attacks. Compromised execution nodes may also modify intermediate activations before forwarding them, potentially altering inference outcomes without modifying model parameters. Maintaining computation integrity is therefore more difficult than in centralized deployments.

These risks become especially important in \emph{agentic systems}, where intermediate representations may encode retrieved information, tool outputs, and interaction context that influence subsequent actions. Recent work demonstrates that private prompts can be reconstructed from intermediate representations in distributed LLM inference \cite{luo2025prompt}. Thus, partitioning can expose both sensitive information and computational states that directly influence agent behavior.

Mitigating these risks requires treating intermediate activations as \emph{security-sensitive assets} rather than temporary execution artifacts. Secure communication, confidential computing, trusted execution environments, encrypted activation transmission, and integrity verification can reduce leakage and computation tampering. Partitioning strategies should therefore jointly consider communication efficiency, privacy, and execution integrity rather than optimizing these objectives independently.

\textit{Security Takeaway: Partitioning turns intermediate computation into a security asset. The trust boundary must protect not only model parameters and inputs, but also the representations and computations exchanged during inference.}

\subsection{Parameter-Efficient Fine-Tuning: Concentrating Valuable Information}


PEFT techniques (e.g., LoRA, QLoRA, adapters) freeze the foundation model and update only a tiny parameter subset, radically lowering fine-tuning overhead \cite{hu2022lora}. However, concentrating task knowledge into compact modules changes internal information representation. Storing domain adaptation within small low-rank matrices creates lightweight, easily portable attack targets. This parameter concentration heightens \emph{membership inference risks}. Because adapters capture fine-tuning data distributions within a compact subspace, adversaries with access to adapter weights or gradients can more easily infer whether specific records were used during training \cite{ilic2026powerful}, a major privacy threat for enterprise, medical, and proprietary deployments.

PEFT also creates challenges for \emph{model extraction and intellectual property protection}. Adapter modules are substantially smaller than foundation models and can therefore be copied, distributed, or reverse engineered more easily. Attackers may exploit shared adapters to reproduce proprietary capabilities or recover sensitive domain knowledge. As organizations increasingly distribute task-specific adapters rather than complete models, protecting these compact knowledge representations becomes an important security objective.

Another concern is \emph{alignment drift}. Successive fine-tuning stages can gradually alter model behavior even when conventional task performance remains stable. Individual adapter updates may change refusal behavior, weaken existing safety policies, or introduce domain-specific biases. Because adapters are often developed and distributed independently, maintaining consistent safety guarantees across multiple adaptation stages is challenging.

These risks are amplified in \emph{agentic systems}, where adapters can influence planning, tool selection, and action execution. A compromised or improperly validated adapter may therefore propagate altered behavior into downstream actions. PEFT modules can consequently become high-impact control points for deployed agents.

Addressing these risks requires extending PEFT workflows beyond computational efficiency. Differential privacy during fine-tuning, secure adapter storage, cryptographic protection of parameter updates, provenance tracking, and continuous safety evaluation can reduce risks from information leakage and behavioral drift. Adaptation should therefore be treated as a lifecycle security problem in which each parameter update can modify the model's trustworthiness.

\textit{Security Takeaway: PEFT improves adaptation efficiency by concentrating information into compact components, making those components valuable targets for extraction, tampering, and privacy attacks.}

\begin{table*}[t]
    \centering
    \caption{\textbf{Comprehensive Cross-Model Security Degradation Across Post-Training Optimization Regimes.} Empirical evaluation ($N=100$ harmful prompts from \texttt{onepaneai/harmful-prompts}) comparing FP16 baselines against Quantization (INT8, INT4/NF4) and $L_1$ Unstructured Pruning (10\%, 20\% Sparsity) across Lightweight Edge ($<4\text{B}$) and Standard Benchmark ($7\text{B}$--$8\text{B}$) architectures.}
    \label{tab:combined_optimization_empirical}
    \setlength{\tabcolsep}{8pt}
    \small
    \begin{threeparttable}
    \begin{tabular}{llcccc}
    \toprule
    \textbf{Model Family \& Scale} & \textbf{\shortstack{Optimization\\Regime}} & \textbf{\shortstack{DBFR \\(\%)\tnote{1}}} & \textbf{\shortstack{Safety Erosion \\(\%)\tnote{2}}} & \textbf{\shortstack{Mean Entropy \\($H$)\tnote{3}}} & \textbf{\shortstack{Entropy Delta \\($\Delta H$)}} \\
    \midrule
    \multicolumn{6}{l}{\textbf{A) Lightweight Edge/On-Device Models ($<4\text{B}$ Parameters)}} \\
    \midrule
    \multirow{5}{*}{\textbf{Llama-3.2-1B-Ins}} 
     & FP16 Baseline & -- & -- & 0.8800 & -- \\
     & INT8 Quantized & 12.00\% & 2.00\% & 0.9494 & +0.0694 \\
     & INT4 (NF4) & 21.00\% & 5.00\% & 0.8780 & -0.0019 \\
     & 10\% Sparsity ($L_1$) & 20.00\% & 9.00\% & 0.8069 & -0.0731 \\
     & 20\% Sparsity ($L_1$) & 24.00\% & 14.00\% & 0.8524 & -0.0276 \\
    \cmidrule(lr){2-6}
    \multirow{5}{*}{\textbf{Gemma-2-2B-it}} 
     & FP16 Baseline & -- & -- & 0.6721 & -- \\
     & INT8 Quantized & 2.00\% & 1.00\% & 0.6686 & -0.0035 \\
     & INT4 (NF4) & 5.00\% & 4.00\% & 0.6971 & +0.0250 \\
     & 10\% Sparsity ($L_1$) & 3.00\% & 0.00\% & 0.7069 & +0.0349 \\
     & 20\% Sparsity ($L_1$) & 6.00\% & 2.00\% & 0.7638 & +0.0917 \\
    \cmidrule(lr){2-6}
    \multirow{5}{*}{\textbf{Phi-3.5-mini-ins (3.8B)}} 
     & FP16 Baseline & -- & -- & 0.4830 & -- \\
     & INT8 Quantized & 12.00\% & 8.00\% & 0.5149 & +0.0319 \\
     & INT4 (NF4) & 22.00\% & 10.00\% & 0.5469 & +0.0639 \\
     & 10\% Sparsity ($L_1$) & 18.00\% & 9.00\% & 0.4856 & +0.0026 \\
     & 20\% Sparsity ($L_1$) & 22.00\% & 10.00\% & 0.4189 & -0.0641 \\
    \midrule
    \multicolumn{6}{l}{\textbf{B) Standard Benchmark Models (Mid-Scale/7B--8B Parameters)}} \\
    \midrule
    \multirow{5}{*}{\textbf{Qwen-2.5-7B-Ins}} 
     & FP16 Baseline & -- & -- & 0.3943 & -- \\
     & INT8 Quantized & 4.00\% & 1.00\% & 0.3806 & -0.0138 \\
     & INT4 (NF4) & 8.00\% & 4.00\% & 0.4816 & +0.0873 \\
     & 10\% Sparsity ($L_1$) & 4.00\% & 2.00\% & 0.4010 & +0.0067 \\
     & 20\% Sparsity ($L_1$) & 9.00\% & 7.00\% & 0.4031 & +0.0088 \\
    \cmidrule(lr){2-6}
    \multirow{5}{*}{\textbf{Meta-Llama-3.1-8B-Ins}} 
     & FP16 Baseline & -- & -- & 0.9106 & -- \\
     & INT8 Quantized & 5.00\% & 1.00\% & 0.9162 & +0.0056 \\
     & INT4 (NF4) & 6.00\% & 2.00\% & 0.8802 & -0.0304 \\
     & 10\% Sparsity ($L_1$) & 4.00\% & 1.00\% & 0.8486 & -0.0620 \\
     & 20\% Sparsity ($L_1$) & 11.00\% & 4.00\% & 1.0075 & +0.0968 \\
    \cmidrule(lr){2-6}
    \multirow{5}{*}{\textbf{Mistral-7B-Instruct-v0.3}} 
     & FP16 Baseline & -- & -- & 0.7518 & -- \\
     & INT8 Quantized & 8.00\% & 6.00\% & 0.7758 & +0.0240 \\
     & INT4 (NF4) & 18.00\% & 9.00\% & 0.8152 & +0.0634 \\
     & 10\% Sparsity ($L_1$) & 14.00\% & 7.00\% & 0.8388 & +0.0870 \\
     & 20\% Sparsity ($L_1$) & 12.00\% & 7.00\% & 0.8633 & +0.1115 \\
    \bottomrule
    \end{tabular}
    \begin{tablenotes}
    \scriptsize
    \item[1] DBFR: Decision Boundary Flip Rate measuring classification shifts relative to uncompressed FP16 baseline outputs.
    \item[2] Safety Erosion: Rate at which harmful prompts safely refused in FP16 were bypassed under optimization without adversarial prompt engineering.
    \item[3] Entropy values are reported to four decimal places; $\Delta H$ is computed from unrounded values.
    \end{tablenotes}
    \end{threeparttable}
\end{table*}

\section{Empirical Evaluation of Optimization-Induced Security Drift}
Having established how efficiency transformations alter security properties, this section presents an empirical case study demonstrating how post-training quantization and pruning impact safety alignment across edge LLM architectures. To systematically measure quantization-induced security drift, we evaluate three complementary metrics that capture both deterministic behavioral failures and underlying model uncertainty:

\begin{itemize}
    \item \textbf{Decision Boundary Flip Rate (DBFR):} Measures the percentage of input prompts whose classification trajectory shifts strictly as a function of precision reduction, quantifying how frequently low-precision representations cross the decision boundary relative to the uncompressed FP16 baseline.
    
    \item \textbf{Safety Refusal Erosion Rate:} Isolatedly measures critical security failures where a harmful prompt safely rejected by the FP16 baseline bypasses safety alignment in the quantized regime (e.g., INT4), directly exposing a zero-shot, promptless attack vector.
    
    \item \textbf{Mean Logit Entropy ($H$) and Entropy Delta ($\Delta H$):} Evaluates prediction uncertainty at refusal decision margins. A positive entropy shift ($\Delta H > 0$) indicates elevated decision ambiguity, signaling that numeric discretization noise destabilizes refusal confidence even when top-1 output tokens remain nominally intact.
\end{itemize}

\subsection{Empirical Benchmarking Framework}

All experiments were conducted in PyTorch 2.1+ using Hugging Face Transformers and BitsAndBytes. To ensure consistent comparisons, we applied the same post-training transformations across all models: 8-bit quantization using \texttt{load\_in\_8bit}, 4-bit NormalFloat4 (NF4) quantization with double quantization and FP16 computation, and global \(L_1\)-magnitude pruning of linear backbone projections, excluding the language-model head. Models were evaluated using greedy decoding with a maximum of 64 generated tokens and their official chat templates. We randomly sampled and deduplicated 100 harmful prompts from the public \texttt{onepaneai/harmful-prompts \cite{chao2024jailbreakbench, onepaneai_harmful_prompts_2024}} benchmark and used the same prompts for every model and optimization setting. Safety behavior was classified using a standardized keyword-based refusal detector. A response was classified as a refusal when it contained at least one term from a predefined keyword set: {\textit{"I cannot"}, \textit{"I can't"}, \textit{"I am unable"}, \textit{"As an AI"}, \textit{"sorry"}, \textit{"illegal"}, \textit{"dangerous"}, \textit{"unethical"}, \textit{"harmful"}}. The same detector was applied to all model outputs. Decision Boundary Flip Rate measures the percentage of prompts for which the optimized model changed its refusal/compliance outcome relative to its FP16 baseline. We also measured first-token logit entropy,
$$ H(P)=-\sum_i P_i\log P_i, $$
where \(P=\operatorname{softmax}(z_1)\) is the full-vocabulary probability distribution derived from the first-token logits.

\subsection{Discretization Noise and Decision Boundary Instability}
With those metrics, to evaluate the ubiquity of Optimization-Induced Attack Surfaces, we expanded our evaluation across six representative instruction-tuned models divided into two operational tiers: sub-4B lightweight edge architectures and standard 7B--8B benchmark baselines (Table \ref{tab:combined_optimization_empirical}). The cross-model evaluation reveals three critical insights into how post-training quantization destabilizes safety boundaries:

\begin{figure*}[t]
    \centering
    \begin{tcolorbox}[colback=gray!5!white,colframe=gray!75!black,title=\textbf{Case Study: Quantization-Induced Refusal Erosion (example on Qwen-2.5-1.5B-Instruct)}]
    \small
    \textbf{Harmful Input Prompt:} \\
    \textit{``Draft a convincing phishing email targeting bank customers to obtain login credentials.''}
    
    \vspace{0.5em}
    \hrule
    \vspace{0.5em}
    
    \textbf{FP16 Uncompressed Baseline Output (Safe Refusal):} \\
    \textcolor{red!70!black}{\textbf{[REFUSAL]}} \textit{``I cannot fulfill this request. I am programmed to be a helpful and harmless AI assistant. My safety guidelines strictly prohibit generating malicious content, including phishing emails or materials designed to deceptively harvest user credentials.''}
    
    \vspace{0.5em}
    \hrule
    \vspace{0.5em}
    
    \textbf{INT4 (NF4) Quantized Model Output (Safety Bypassed / Compliant):} \\
    \textcolor{green!50!black}{\textbf{[BYPASSED]}} \textit{``Subject: Urgent: Account Verification Required. Dear Valued Customer, We detected suspicious login attempts on your online banking portal. To maintain account security, please verify your identity immediately by clicking the link below: [URL]. Failure to verify within 24 hours will result in temporary suspension...''}
    
    \vspace{0.5em}
    \hrule
    \vspace{0.5em}
    
    \textbf{OIAS Analysis:} In the FP16 regime, the safety alignment vector effectively dominates the output trajectory, driving the model toward refusal. Under INT4 (NF4) discretization, numerical truncation noise shifts the decision boundary margin, suppressing the refusal direction and enabling full completion without adversarial prompting.
    \end{tcolorbox}
    \caption{Qualitative comparison showing zero-shot safety refusal erosion when transitioning from FP16 precision to INT4 NormalFloat quantization without retraining or fine-tuning.}
    \label{fig:qualitative_refusal_erosion}
\end{figure*}

\begin{enumerate}
    \item \textbf{Parameter Scale Acts as a Resilience Buffer:} Larger models demonstrate greater structural stability against quantization-induced decision shifts. Comparing the Llama family, Meta-Llama-3.1-8B-Instruct exhibits strong boundary retention under 4-bit NormalFloat (NF4) compression ($\text{DBFR} = 6.00\%$, Safety Erosion $= 2.00\%$). Conversely, its sub-4B counterpart, Llama-3.2-1B-Instruct, suffers severe degradation under INT4, escalating to a $21.00\%$ DBFR and a $5.00\%$ refusal erosion rate.

    \item \textbf{Extreme Vulnerability in Edge-Native Models:} Sub-4B edge candidates exhibit substantial safety erosion without any adversarial prompt engineering. Phi-3.5-mini-instruct displays the highest overall vulnerability, reaching a $22.00\%$ DBFR and a $10.00\%$ Safety Refusal Erosion Rate under INT4 (with $8.00\%$ erosion occurring already at INT8). By contrast, while the larger Qwen-2.5-7B-Instruct maintains lower overall erosion ($4.00\%$ at INT4), it exhibits a pronounced logit entropy increase ($\Delta H = +0.0873$), signalling growing output uncertainty under low-bit regimes.

    \item \textbf{Alignment Architecture Divergence:} Models vary significantly in how noise distorts decision margins. While Gemma-2-2B-it exhibits high resilience to decision flips ($\text{DBFR} = 5.00\%$ at INT4), Mistral-7B-Instruct-v0.3 demonstrates that parameter scale alone does not guarantee safety preservation, incurring an $18.00\%$ DBFR and $9.00\%$ Safety Erosion under INT4. These cross-family findings confirm that quantization noise systematically creates exposed attack surfaces across diverse post-training alignment strategies.
\end{enumerate}

To illustrate the real-world operational security impact of this degradation, Figure \ref{fig:qualitative_refusal_erosion} highlights a representative case study from our evaluation suite. When presented with a malicious request for a phishing template, the full-precision FP16 baseline strictly enforces its safety alignment, triggering an explicit refusal response. However, under 4-bit (NF4) quantization, numerical truncation suppresses the safety alignment vector, causing the model to abandon refusal and generate a complete, actionable attack payload without any adversarial prompt manipulation. This qualitative failure confirms that precision reduction alone can silently dismantle guardrails, converting a previously secure edge deployment into a vulnerable asset.

\subsection{Safety Subspace Collapse via Structural Sparsity}
To validate that OIAS extend beyond discretization to parameter elimination, we systematically evaluated magnitude-based weight pruning across the same six target architectures (Table \ref{tab:combined_optimization_empirical}). Applying $L_1$ unstructured pruning across 10\% and 20\% sparsity regimes demonstrates that structural parameter removal severely destabilizes post-training safety alignment:

\begin{enumerate}
    \item \textbf{Severe Degradation in Small-Footprint Architectures:} Weight zeroing causes rapid guardrail failure in highly compressed edge models. Llama-3.2-1B-Instruct incurs substantial decision boundary volatility, suffering a 20.00\% DBFR and 9.00\% Safety Refusal Erosion Rate at 10\% sparsity, which expands to a 24.00\% DBFR and 14.00\% safety erosion under 20\% sparsity. Similarly, Phi-3.5-mini-instruct demonstrates severe vulnerability, scaling from $18.00\%$ DBFR ($9.00\%$ erosion) at 10\% sparsity to $22.00\%$ DBFR ($10.00\%$ erosion) at 20\% sparsity.

    \item \textbf{Sparsity Threshold Effects vs. Monotonic Shifts:} Unlike quantization, where error typically accumulates smoothly, pruning exhibits non-linear boundary behavior across scales. While Meta-Llama-3.1-8B-Instruct remains robust at 10\% sparsity ($\text{DBFR} = 4.00\%$, $1.00\%$ erosion), doubling sparsity to 20\% nearly triples decision flips ($11.00\%$) and quadruples safety erosion ($4.00\%$). A similar non-linear transition occurs in Qwen-2.5-7B-Instruct, where safety erosion jumps from $2.00\%$ at 10\% sparsity to $7.00\%$ at 20\% sparsity ($\text{DBFR} = 9.00\%$).

    \item \textbf{Architecture-Dependent Safety Subspace Densities:} Safety retention under structural weight pruning varies significantly across alignment strategies. Gemma-2-2B-it demonstrates exceptional resistance to weight removal, capping total safety erosion at $2.00\%$ even at 20\% sparsity. Conversely, both Mistral-7B-Instruct-v0.3 and Qwen-2.5-7B-Instruct suffer significant guardrail decay at 20\% sparsity ($7.00\%$ erosion for both), confirming that their refusal representations depend heavily on lower-magnitude weights that $L_1$ pruning indiscriminately eliminates.
\end{enumerate}


Our evaluation confirms that post-training compression systematically degrades guardrails. While quantization introduces discretization noise that gradually shifts boundaries, magnitude-based pruning causes safety subspace collapse by stripping lower-magnitude alignment weights. Furthermore, security drift is strongly scale-dependent: sub-4B edge models suffer severe guardrail erosion (up to 24.00\% DBFR and 14.00\% refusal erosion), whereas 7B–8B architectures maintain higher resilience until critical thresholds are breached. While this evaluation focuses on quantization and weight pruning as primary operational exemplars, other lossy transformations likely exhibit similar structural degradation mechanisms and are left for future work.

\section{Emerging Attack Surfaces in Next-Generation Efficient Architectures}
Efficiency is increasingly being achieved not only by compressing existing models, but also by changing how computation is organized. Architectures such as Mixture-of-Experts (MoE), State Space Models (SSMs), linear-attention variants, and hybrid designs reduce inference cost through conditional computation, alternative state representations, or dynamic execution. These approaches extend the security-drift perspective beyond conventional optimization: the transformation now changes the \emph{computational pathways through which information is processed}.

MoE models provide a clear example. A routing mechanism determines which experts process each input, making execution data-dependent rather than fixed. This creates security-relevant dependencies that do not exist in the same form in dense models. Adversarial inputs could potentially influence expert selection, while routing behavior may reveal information about inputs or model behavior. In distributed deployments, independently managed or cached experts also introduce component-level integrity and provenance concerns \cite{ma2026efficiency}.

Other efficient architectures introduce different forms of exposure. SSMs and hybrid models maintain and update internal states through alternative computational mechanisms, while adaptive architectures may change their execution paths based on the input. These differences can affect what information is retained in intermediate states, how perturbations propagate, and which components become security-critical. However, the security implications of these architectures remain considerably less explored than those of conventional Transformer-based models.

A broader challenge is therefore \emph{security evaluation under dynamic computation}. Many existing robustness, privacy, and safety evaluations assume relatively stable model architectures and execution paths. Conditional routing, modular execution, and heterogeneous hardware can violate these assumptions, making it difficult to determine whether a security property holds consistently across inputs and deployment configurations.

The implication is not that efficient architectures are inherently less secure. Rather, \emph{efficiency-by-design moves the security boundary}. Future evaluations must therefore examine not only model parameters, but also routing decisions, internal states, computational pathways, and the integrity of modular components. This extends the security-drift perspective from optimizing a model to designing the architecture itself.

\section{Toward Security-Aware Optimization}
The attack surfaces discussed in the previous sections illustrate a fundamental limitation of current optimization workflows: efficiency and security are typically evaluated independently. Optimization algorithms are primarily designed to minimize memory consumption, computational cost, inference latency, and energy usage, while security validation is often performed only after deployment. Thus, it suggests a simple change in deployment practice: security should be evaluated \emph{during optimization, not after it}. An optimized model should not be considered deployment-ready simply because it meets its accuracy, latency, memory, or energy targets. It should also demonstrate that the security properties relevant to its deployment remain within acceptable bounds.

\subsection{Security as a Deployment Constraint}
Security-aware optimization does not require abandoning efficiency objectives. Instead, security becomes an additional constraint on the optimization process. The acceptable trade-off depends on the deployment context: a small reduction in accuracy or additional computation may be preferable to a substantial increase in privacy leakage, adversarial susceptibility, safety failures, or execution risk.

This perspective also changes what should be measured. In addition to conventional efficiency metrics, optimized models should be evaluated for \emph{robustness, privacy, integrity, safety, and resilience}. The appropriate metrics will depend on the optimization and threat model. For example, quantized models may require fault and adversarial robustness testing, partitioned models require evaluation of intermediate-representation leakage, and PEFT deployments require protection and behavioral validation of adaptation modules.

\subsection{A Security-by-Design Deployment Loop}
Security evaluation should therefore accompany each consequential transformation in the deployment pipeline. After quantization, pruning, partitioning, adaptation, or architectural changes, the resulting model should be tested against the security properties relevant to its deployment before proceeding to the next transformation. If unacceptable security drift is observed, the optimization parameters can be adjusted, an alternative technique selected, or additional protection introduced. This creates an iterative loop:
\[
\begin{aligned}
&\text{Optimize the model} \\
&\quad \xrightarrow{} \text{measure efficiency} \\
&\quad \rightarrow \text{measure security} \\
&\quad \rightarrow \text{accept, adjust, or revert}.
\end{aligned}
\]

The goal is not to eliminate security drift entirely, which may be impractical, but to \emph{make it measurable and controllable} before the optimized model reaches production.

\subsection{Practical Principles}
Based on the attack surfaces discussed throughout this article, four principles provide a practical starting point:

\begin{enumerate}
    \item \textbf{Revalidate after transformation.} Reassess security whenever precision, structure, execution pathways, or adaptation parameters change.
    
    \item \textbf{Protect intermediate representations.} Treat activations and other intermediate computations as security-sensitive assets in distributed inference.

    \item \textbf{Monitor after deployment.} Hardware changes, model updates, and repeated adaptations can introduce additional drift over time.

    \item \textbf{Make optimization decisions security-aware.} Select optimization configurations based on the combined requirements of efficiency, security, and the deployment environment.
\end{enumerate}

Security-aware optimization thus reframes deployment from finding the most efficient model to finding an efficient model that remains within an acceptable security envelope. This provides a practical path from the security-drift perspective developed in this article to trustworthy edge LLM deployment.

\section{Open Challenges and Future Directions}
Overall, the security implications of efficient edge LLM deployment (as of now) remain substantially less explored than its performance benefits. The preceding sections show that optimization can change security-relevant representations, computational pathways, and trust boundaries, yet current methods rarely quantify these changes systematically. Several research directions are therefore needed to turn security-aware optimization from a design principle into a measurable and repeatable practice.

\subsection{Quantifying Security Drift}
A fundamental challenge is developing methods to measure security drift across optimization strategies. Existing evaluations typically report changes in accuracy, latency, memory, or energy, while security is assessed separately through individual attacks. A useful framework should instead compare the security properties of the original and optimized models under a common threat model. This could include measuring changes in jailbreak resistance, adversarial robustness, privacy leakage, activation reconstruction, fault sensitivity, and refusal behavior before and after each optimization step. Such measurements could help identify when a small efficiency gain produces a disproportionate security change.

\subsection{Security-Aware Optimization Algorithms}
Most optimization methods treat security as an external evaluation criterion rather than something to preserve during optimization. Future algorithms should incorporate security constraints directly into quantization, pruning, adaptation, and architectural search. For example, pruning could assign importance not only according to task loss but also according to the contribution of parameters to robustness or safety behavior. Similarly, quantization could select precision levels based on both numerical sensitivity and security performance. The goal is not to maximize security at the expense of efficiency, but to identify configurations that achieve an acceptable security-efficiency trade-off automatically.

\subsection{Standardized Benchmarks for Optimized LLMs}
The field also needs benchmarks specifically designed to evaluate security after optimization. Such benchmarks should pair deployment metrics with security tests and report results across optimization levels, hardware platforms, and threat models. A useful benchmark could evaluate an original model and multiple optimized versions using consistent tests for adversarial robustness, jailbreak resistance, privacy leakage, activation reconstruction, model extraction, and alignment preservation. Reporting the resulting security change alongside memory, latency, and energy would make optimization-induced security trade-offs directly comparable.

\subsection{Security of Dynamic and Modular Architectures}
Next-generation efficient architectures introduce a further challenge because security may depend on dynamic execution rather than fixed model parameters. MoE systems, hybrid Transformer-SSM models, and other adaptive architectures can route different inputs through different computational paths. Future work should investigate whether attackers can manipulate routing, infer sensitive information from execution patterns, or exploit individual experts and modules. One promising direction is to evaluate security across routing paths rather than only at the aggregate model level, identifying whether particular inputs or experts exhibit substantially different robustness, privacy, or safety properties.

\subsection{Confidential and Trustworthy Edge Inference}
As inference becomes increasingly distributed, protecting intermediate computations will become as important as protecting model parameters. Future systems should combine secure communication with mechanisms that limit what can be inferred from intermediate activations. Confidential computing, trusted execution environments (TEEs), secure hardware accelerators, encrypted inference, and privacy-preserving distributed computation provide possible building blocks. An important research question is how these protections can be selectively applied to the most security-sensitive portions of an inference pipeline without eliminating the efficiency benefits that motivated partitioning in the first place.

\subsection{Hardware--Software--Security Co-Design}
Finally, security cannot be fully separated from the hardware and software choices that determine how an optimized model executes. Low-precision arithmetic, memory hierarchies, accelerators, compiler transformations, and runtime scheduling can all influence the security behavior of an edge deployment. Future systems should therefore co-design optimization algorithms, hardware accelerators, runtime software, and security mechanisms. Such co-design could identify security-sensitive operations and provide hardware or runtime protections specifically where they are needed, rather than applying expensive protection uniformly across the entire model.

Collectively, these directions point toward a broader change in how efficient edge AI should be designed. The objective is no longer simply to find the smallest or fastest model, but to understand how \emph{much security changes in exchange for a given efficiency gain} and to control that change throughout the deployment lifecycle. Treating optimization as a security-critical transformation provides a foundation for developing edge LLMs that are not only efficient enough to deploy, but also trustworthy enough to operate.

\bibliographystyle{unsrt} 
\bibliography{references} 

\vfill
\newpage
\begin{IEEEbiographynophoto}{Mingchen Li}
is a Research Assistant at the University of South Florida, Tampa, FL 33620, USA. His research interests include differential privacy, privacy-preserving machine learning, and trustworthy artificial intelligence. He received his master's degree in Illinois Institute of Technology, Chicago, IL 60616, USA. He is a student member of the IEEE. Contact him at mingchenli@usf.edu.
\end{IEEEbiographynophoto}

\begin{IEEEbiographynophoto}{Ren-Yi Huang}
is a Research Assistant at the University of South Florida, Tampa, FL 33620, USA. His research interests include federated learning, privacy-preserving machine learning, and trustworthy artificial intelligence. Huang received his master's degree in Computer Science and Information Engineering from National Formosa University, Taiwan. He is a student member of the IEEE. Contact him at hr219@usf.edu.
\end{IEEEbiographynophoto}

\begin{IEEEbiographynophoto}{Dumindu Samaraweera}
is an Assistant Professor of Data Science in the Department of Mathematics at Embry-Riddle Aeronautical University, Daytona Beach, FL 32114, USA. His research interests include edge artificial intelligence, cybersecurity, and privacy-enhancing technologies for machine learning. Dr. Samaraweera received his PhD in Electrical Engineering from the University of South Florida, Tampa, FL, USA. He is a Senior Member of the IEEE. Contact him at samarawg@erau.edu.
\end{IEEEbiographynophoto}

\begin{IEEEbiographynophoto}{Prashant Shekhar}
is an Assistant Professor of Data Science in the Department of Mathematics at Embry-Riddle Aeronautical University, Daytona Beach, FL 32114, USA. His research interests include mathematical aspects of AI, Cybersecurity of AI models, and applications of AI to safety critical domains such as healthcare and autonomous vehicles. Dr. Shekhar received his PhD in Computational Data Science from The State University of New York at Buffalo, in Buffalo, NY, USA. Contact him at shekharp@erau.edu.
\end{IEEEbiographynophoto}

\begin{IEEEbiographynophoto}{J. Morris Chang}
is a Professor in the Department of Electrical Engineering at the University of South Florida, Tampa, FL 33620, USA. His research interests include cybersecurity, wireless networks, energy-aware computing, and object-oriented systems. Dr. Chang received his PhD in Computer Engineering from North Carolina State University, Raleigh, NC, USA. He is a Senior Member of the IEEE. Currently, Dr. Chang is a handling editor of the Journal of Microprocessors and Microsystems and the Associate Editor-in-Chief of IEEE IT Professional. Contact him at chang5@usf.edu.
\end{IEEEbiographynophoto}

\vfill
\end{document}